\documentclass[aps,prx,reprint,longbibliography,nofootinbib,superscriptaddress,floatfix]{revtex4-2}

\usepackage{slashed}
\usepackage{verbatim}
\usepackage[T1]{fontenc}
\usepackage{mathbbol}
\usepackage[dvipsnames]{xcolor}
\usepackage{orcidlink}
\usepackage[normalem]{ulem}
\usepackage[english]{babel}
\usepackage{lipsum}
\usepackage{adjustbox}
\usepackage{physics}
\usepackage{dcolumn}
\usepackage{tensor}
\usepackage{comment}
\usepackage{graphicx}
\usepackage{color,overpic,mathtools}
\usepackage{amsthm,amsmath,amssymb,mathrsfs}
\usepackage{braket,bm,bbm,setspace}
\usepackage{cancel}
\usepackage{float}
\usepackage{xargs}
\definecolor{myred}{RGB}{179, 27, 27}
\definecolor{teal}{RGB}{0, 128, 128}
\usepackage{hyperref}
\usepackage{rotating}
\hypersetup{
    colorlinks=true,
    linkcolor=teal, 
    citecolor=teal, 
    urlcolor=teal  
 }

\newcommand{\be}{\begin{equation}}
\newcommand{\ee}{\end{equation}}
\newcommand{\beq}{\begin{eqnarray}}
\newcommand{\eeq}{\end{eqnarray}}

\begin{document}
%
\title{Overdamped quasibound states inside a Schwarzschild black hole}
\author{Jeff Steinhauer\orcidlink{0000-0002-0757-3624}}
\affiliation{Department of Physics, Technion -- Israel Institute of Technology, Haifa 3200003, Israel}
\affiliation{CENTRA, Departamento de Física, Instituto Superior Técnico – IST, Universidade de Lisboa – UL, Avenida Rovisco Pais 1, 1049-001 Lisboa, Portugal}
\author{Kyriakos Destounis\orcidlink{0000-0003-2997-088X}}
\affiliation{CENTRA, Departamento de Física, Instituto Superior Técnico – IST, Universidade de Lisboa – UL, Avenida Rovisco Pais 1, 1049-001 Lisboa, Portugal}
\author{Richard Brito\orcidlink{0000-0002-7807-3053}}
\affiliation{CENTRA, Departamento de Física, Instituto Superior Técnico – IST, Universidade de Lisboa – UL, Avenida Rovisco Pais 1, 1049-001 Lisboa, Portugal}

\begin{abstract}
Schwarzschild black-hole interiors, bounded by event horizons and terminated by spacelike singularities, are regions where all physical observers are inevitably destroyed. In the geometric optics approximation, waves follow null geodesics to the singularity. However, outside the geometric optics regime, the behavior of wave propagation can be rich and nuanced, even in such extreme habitats. In this work, we show that axial gravitational perturbations in the interior of a Schwarzschild black hole can form overdamped (non-oscillatory) quasibound states that decay before reaching the singularity. Using Kruskal-Szekeres coordinates to avoid coordinate ambiguities, we identify these modes and analyze their eigenfunctions. Contrary to earlier claims, we find that the Regge-Wheeler master function of these modes have non-zero amplitude at the future event horizon but decay before interacting with the singularity. We consider observations of the modes along timelike geodesics. This work suggests that certain gravitational fluctuations can hover transiently within the black-hole interior, challenging common assumptions about wave behavior in uncharted and extreme regions of spacetime.
\end{abstract}

\maketitle

\section{Introduction}

A foundational feature of black holes (BHs) is a demarcation hypersurface that causally disconnects the BH interior from null infinity; \emph{whatever happens in the interior stays in the interior}. The external region is, of course, paramount for gravitational-wave (GW) astrophysics \cite{Schutz:1999xj,Bishop:2021rye,Bailes:2021tot} since the explicit information of BH geometries can be radiated away to infinity, and thus reach our detectors through GWs. In fact, the whole field of BH astrophysics is unavoidably linked to global quantities of various BH merger events that emit GWs as the coalescence takes place \cite{Buonanno:2006ui,Buonanno:2007sv,Pan:2011gk,Ghosh:2017gfp,Pompili:2025cdc}, till a final remnant is formed. It, therefore, seems that we can perform observations without having to worry about what occurs \emph{behind the obscurity} of event horizons, since the interior is causally-disconnected with the exterior.

Nevertheless, the interior of a BH is an indispensable part of its causal structure \cite{Poisson:1990eh}, thus in order to understand BHs in a holistic way we also have to examine their interior. A plethora of studies have been conducted in order to better comprehend the elusiveness of BH interiors. One of the most fundamental aspects of a BH interior is the curvature singularity \cite{Penrose:1964wq,Hawking:1971vc,Ellis:1977pj,Wheeler:2022zuh}. The singularity itself cannot be unequivocally understood since the laws of Physics, as we know them, break down there\footnote{See though \cite{Kerr:2023rpn} for an interesting discussion on the existence of singularities.}. Mathematically, we can classify singularities \cite{Tipler:1977zza,Nolan:1999tw}, as strong \cite{Ori:1991zz,Brady:1995ni,Brady:1998ht,Hod:1998gy,Herman:1992uf,Dafermos:2003vim,Dafermos:2012np,Luk:2022rgs,Luk:2015pay,Luk:2015qja,Dafermos:2015bzz,Kehle:2018upl,Kehle:2018zws,Kehle:2021ufl} or weak \cite{Ori:1992zz,Ori:1995nj,Ori:1998az,Burko:1997fc,Burko:1997zy,Burko:1998jz,Burko:1999zv,Luk:2013cqa,Gajic:2015csa,Gajic:2015hyu,Gajic:2017bti,Hintz:2015jkj,Hintz:2015koq}, where the interpretation of strength stems from the effect that a singularity imposes on infalling observers, i.e. how much regularity these observers maintain upon encountering a singularity \cite{Dafermos:2003wr}. In fact, it is the strength of a singularity that defines the violation of the Strong Cosmic Censorship conjecture \cite{Christodoulou:1987vv,Christodoulou:2008nj,Luk:2017jxq,Kehle:2020zfg,Kehle:2021jsp}, which states that potential extensions of spacetime beyond the boundary of maximal globally-hyperbolic development of initial data, i.e. the Cauchy horizon, should have square-integrable Christoffel symbols, in order to form, at least, weak solutions to the field equations in a local vicinity of the Cauchy horizon \cite{Christodoulou:2008nj}. This means that even though singularities may persist inside BHs, they might be weak enough to allow for extensions beyond it. The strength of singularities has also been associated with the quasinormal modes (QNMs) of BH exteriors in cosmological BH spacetimes \cite{Cardoso:2017soq,Cardoso:2018nvb,Mo:2018nnu,Dias:2018ufh, Dias:2018ynt,Ge:2018vjq,Destounis:2018qnb,Rahman:2018oso,Liu:2019lon,Liu:2019rbq,Destounis:2019omd,Guo:2019tjy,Destounis:2020yav,Luna:2019olw,Casals:2020uxa,Konoplya:2022kld}. 

Thus, studying the dynamical behavior of BHs is extremely important in order to achieve a robust understanding of how perturbations evolve in time and space. In BH exteriors, linear and non-linear excitations of asymptotically-flat and de Sitter geometries are bounded and decay with time (though grow with distance) \cite{Dafermos:2010hb,Dafermos:2014cua,Dafermos:2014jwa,Dafermos:2016uzj,Dafermos:2017yrz,Dafermos:2021cbw,Hintz:2013kbc,Hintz:2013lka,Hintz:2014rsa,Hintz:2016gwb,Hintz:2016jak,Hintz:2020roc,Hintz:2023aus}, according to their spectral properties described by QNMs \cite{Press:1971wr,Davis:1972ud,Kokkotas:1999bd,Berti:2009kk,Dreyer:2003bv,Berti:2009kk,Hughes:2019zmt,Destounis:2023ruj,Brito:2018rfr,Berti:2025hly,Jaramillo:2020tuu,Jaramillo:2022kuv,Destounis:2021lum,Ghosh:2021mrv,Destounis:2023nmb,Sarkar:2023rhp,Boyanov:2022ark,Boyanov:2023qqf,Cheung:2021bol,Boyanov:2022ark,Boyanov:2024fgc,Cai:2025irl,Rosato:2024arw,Cardoso:2021wlq,Cardoso:2022whc,Pezzella:2024tkf,Dyson:2025dlj}. Quasibound states (QBSs) \cite{Zouros:1979iw,Detweiler:1980uk,Dolan:2007mj,Dolan:2015eua,Hod:2012zza,Hod:2017gvn,Destounis:2019hca,Huang:2020pga,Senjaya:2024uqg,Mascher:2022pku,Vieira:2021xqw,Vieira:2023ylz,Vieira:2023ylz,Vieira:2025ljl,Vieira:2025gcy} of massive fields behave differently from QNMs due to different boundary conditions imposed at future null infinity. In fact, QBSs of massive fields, can grow or decay with time (and decay with distance). In BH interiors though, the dynamics of perturbations becomes more subtle and intricate. Actually, even though a potential spectrum in the interior of a BH might exist, we practically utilize it for purely theoretical reasons \cite{Miguel:2020uln}, in order to recognize the importance of singularities, realize the limitations of classical General Relativity (GR), and deduce models of quantum gravity \cite{Ashtekar:1986yd,Ashtekar:2021kfp,Rovelli:1987df,Rovelli:1989za,Kiritsis:2019npv,Susskind:1993ws,Skenderis:2008qn,Russo:1994ev,Mathur:2005zp,Henneaux:2007ej,Berkooz:2007nm} that can partially or completely erase singularities from BH interiors. All studies discussed above regarding the stability of BH interiors and Cauchy horizons in charged and rotating spacetimes share a common starting point; they begin with an initial value problem that is set in the exterior and evolves the initial data beyond the event horizon, and into the BH interior, where singularities lurk. A very important and systematic question is then posed: \emph{What is the behavior of perturbations that begin in the BH interior? Do they simply plunge into the singularity, or can they exhibit a more complicated behavior?}

In this work, after analyzing the axial gravitational perturbations of the Schwarzschild BH interior, by imposing boundary conditions that are mathematically consistent with the geometry and the perturbation equations, we find purely-damped (non-oscillatory) modes that are localized solely in the internal region of the BH. We name these solutions as \emph{overdamped quasibound states} (OQBSs). These, otherwise, regular solutions are sourced through the emergence of a negative well for axial gravitational perturbations in the BH interior, that diverges to $+\infty$ close to the singularity. The respective wavefunctions of these OQBSs encounter the future event horizon but decay before encountering the singularity, or the past horizon, in the same manner as a QBS wavefunction would behave for $r\rightarrow\infty$ in the BH exterior, where $r$ is the usual radial coordinate in Schwarzschild coordinates. In order to avoid certain ambiguities regarding the nature of Schwarzschild coordinates inside the BH, and bypass the coordinate singularity at the event horizon, we have achieved this analysis through the use of Kruskal–Szekeres coordinates \cite{Kruskal:1959vx,Szekeres:1960gm,Israel:1966zz}, which have not yet been used for such a spectral problem. Our findings suggest that the OQBSs are regular everywhere in the Schwarzschild interior, hover transiently and eventually decay with respect to proper time, before encountering the curvature singularity.

The spectrum of Schwarzschild BH interiors was also studied in Ref. \cite{Firouzjahi:2018drr} for scalar, axial and polar gravitational perturbations, but it was found that the internal modes did not encounter the event horizon, in contrast to our results. Ref. \cite{Firouzjahi:2018drr} also connected the internal modes with asymptotic QNMs. Another study of the interior dynamics of a BH was performed in Ref. \cite{Fiziev:2006tx}, though the ``spectral problem'' solved in the interior used an inverted potential technique, as in Refs. \cite{Ferrari:1984ozr,Ferrari:1984zz,Volkel:2025lhe}, in order to find analytical bound states, through confluent Heun functions. Therefore, the bound state problem in \cite{Fiziev:2006tx} is fundamentally different from the one that we present here. In what follows we assume the geometrized unit system, i.e., $G=c=1$.

\section{Schwarzschild black holes and perturbations}\label{SecII}

The Schwarzschild geometry is the unique asymptotically flat BH solution to the vacuum Einstein field equations in spherical symmetry \cite{Hawking:1973uf}. The solution describes a BH with mass $M$, while its causal structure exhibits a null hypersurface at $r=2M$ that causally disconnects information from future null infinity when $r<2M$. On the other hand, when $r>2M$ information can reach future null infinity. Thus, the hypersurface at $r=2M$ separates the interior of the Schwarzschild BH from its exterior. Finally, at the BH interior there is a spacelike curvature singularity at $r=0$, where the Kretschmann scalar blows up. No observer can survive a spacelike singularity due to its infinite tidal forces and tidal deformations inflicted on it.

A Schwarzschild BH spacetime metric can be written using the typical Schwarzschild coordinates $(t,r,\theta,\varphi)$ as \cite{Hawking:1973uf}
\begin{equation}\label{Schwarzschild_metric}
    ds^2=-f(r)\,dt^2+f^{-1}(r)\,dr^2+r^2\left(d\theta^2+\sin^2\theta\,d\varphi^2\right),
\end{equation}
where
\begin{equation}
    f(r)=1-\frac{2M}{r},
\end{equation}
is the lapse function. Even though the Schwarzschild coordinates are not ideal to study the interior and exterior without coordinate ambiguities, it is predominantly used in the literature to study the Schwarzschild exterior. Alternatives are the ingoing and outgoing Eddington-Finkelstein, double null and Kruskal-Szekeres coordinates, among others \cite{Hawking:1973uf}.

The study of small perturbations in Schwarzschild BHs has a long history \cite{Chandrasekhar:1975zza,Chandrasekhar:1985kt}. Essentially, perturbations to the metric propagate as GWs on a fixed background; in our case the Schwarzschild background is described by the line element \eqref{Schwarzschild_metric}. Under the assumption of harmonic dependence in the Schwarzschild coordinate $t$ and by using the tensor spherical harmonics for the angular part's eigenvalues, i.e. the spherical-harmonic index $\ell$, axial gravitational perturbations can be written in terms of a gauge-invariant master function $\psi(r)$ that satisfies the Regge-Wheeler equation \cite{Regge:1957td,Martel:2005ir} 
\begin{equation}\label{RW_equation}
    \frac{\partial^2\psi}{\partial r_*^2}+\left(\omega^2-V_\textrm{RW}\right)\psi=0.
\end{equation}
Here, $\omega$ is the propagation frequency of the axial GW. The tortoise coordinate in Eq. \eqref{RW_equation} is defined as
\beq
r_*=r+2M\,\mathrm{ln}(r/2M-1),\;\;\;\; (r>2M), \label{eq1}
\eeq
which satisfies
\beq
dr_*=dr/(1-2M/r). \label{eq2}
\eeq

\begin{figure}[t]
    \centering
    \includegraphics[width=\columnwidth]{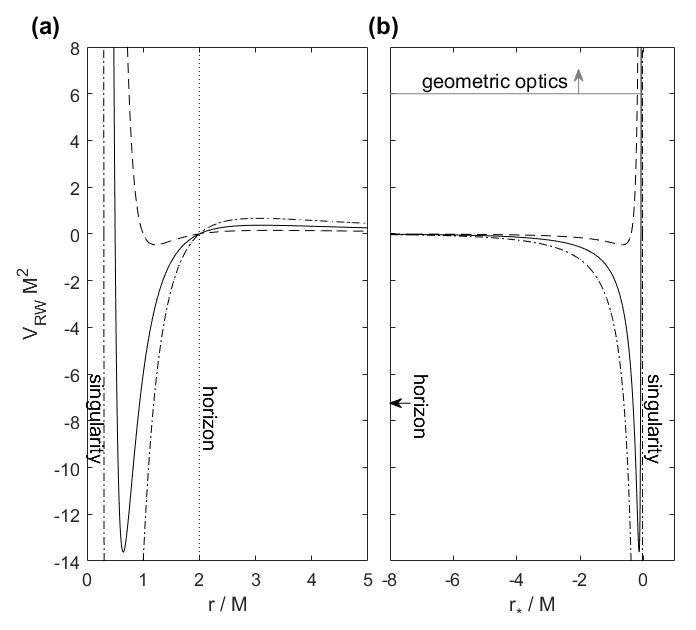}
    \caption{The continuation of the Regge-Wheeler effective potential $V_\mathrm{RW}$ inside the Schwarzschild black hole. The dashed, solid and dash-dotted curves correspond to $\ell = 2,\, 3$ and $4$, respectively. (a) $V_\mathrm{RW}$ as a function of $r$. (b) $V_\mathrm{RW}$ as a function of $r_*$. The horizon is at $r_*=-\infty$. The horizontal ``geometric optics'' line indicates a large value of $\omega$ (not to scale).}
    \label{fig:potential}
\end{figure}

The Regge-Wheeler potential for axial gravitational perturbations is given by \cite{Regge:1957td}
\beq
V_\mathrm{RW}=f(r)\left[\frac{\ell(\ell+1)}{r^2}-\frac{6M}{r^3} \right]. \label{eq4}
\eeq
The study of perturbations and resulting spectra in the external region of BHs is an extremely fine-studied field of BH physics. Depending on the boundary conditions imposed in the perturbations we can obtain QNMs \cite{Kokkotas:1999bd,Berti:2009kk} or QBSs (for massive scalar, Dirac, vector and tensor field perturbations) \cite{Zouros:1979iw,Dolan:2007mj,Dolan:2015eua,Rosa:2011my,Brito:2013wya,Babichev:2015zub,Brito:2015oca} that differ through the boundary condition at infinity; the former have ingoing waves at the event horizon and outgoing waves at infinity, while the latter decay exponentially at spatial infinity.

Although there are significantly fewer studies regarding the BH interior, mostly due to the fact that no observables can be produced and propagate to the external region where measurements are performed, the internal region of a BH is an inseparable part of spacetime. Therefore, its structure is equally important and fundamentally fascinating in order to completely comprehend GR and its limitations. In fact, according to Penrose's weak cosmic censorship conjecture \cite{Penrose:1969pc,Wald:1997wa}, event horizons are formed through gravitational collapse of matter \cite{Penrose:1964wq} in order to hide spacetime singularities in BH interiors where the physical laws break down. Thus, the study of perturbations and how they affect BH interiors and lead to magnificent effects, such as the creation of mass-inflation singularities in charged and rotating black holes \cite{Ori:1991zz}, is in itself rather compelling.

Inside the BH, the tortoise coordinate becomes
\beq
r_*=r+2M\, \mathrm{ln}(1-r/2M),\;\;\;\; (r<2M), \label{eq5}
\eeq
i.e. the sign changes inside the logarithm. In turn, in the interior we have $-\infty<r_*<0$, where the singularity $r=0$ is mapped to $r_*=0$ and the event horizon $r=2M$ is mapped to $r_*\to -\infty$. Nevertheless, the Regge-Wheeler equation utilizes derivatives of the tortoise coordinate that still satisfy Eq. \eqref{eq2} inside the BH. In other words, the expression for $r_*$ is different inside and outside of the BH, but the expression for $dr_*$ is the same everywhere in the domain of $r_*$. Thus, Eqs. \eqref{RW_equation} and \eqref{eq4} are also valid in the Schwarzschild BH interior. We will take advantage of this, and attempt to set up a proper spectral problem in the interior of a Schwarzschild BH, that originates directly in the interior, i.e. initial data are given inside the BH for the evolution of small perturbations. Our analysis is directly based on the Regge–Wheeler equation, which describes axial gravitational perturbations in the Schwarzschild geometry. While inside the event horizon $r$ and $r_*$ are timelike, this does not affect the validity of the equation itself, which remains mathematically well-defined in both the exterior and interior of the BH spacetime. In what follows, we retain the conventional terminology of the ``tortoise coordinate'', in order to remain consistent with the literature. By choosing appropriate boundary conditions in the interior we find the spectrum of OQBSs. This approach was used in previous work to compute QNMs in the interior of other BH geometries~\cite{Miguel:2020uln,Matzner:1980}. However, even though the initial calculations are performed in Schwarzschild coordinates, we ultimately use the Kruskal-Szekeres coordinate system in order to interpret the physical character of these modes.

\section{Overdamped quasibound states inside a black hole}

Fig. \ref{fig:potential}(a) demonstrates the effective potential Eq. \eqref{eq4} of axial gravitational perturbations both inside and outside the Schwarzschild BH. There is a weak maximum in the exterior of the BH ($r > 2M$) which results in the well-known excitations of the light ring and the resulting QNMs \cite{Chandrasekhar:1975zza}. An effective potential minimum appears in the BH interior, between the event horizon and the singularity, in both Fig. \ref{fig:potential}(a) which shows the potential with respect to $r$, as well as in Fig. \ref{fig:potential}(b) that shows the potential as a function of the ``tortoise coordinate'' $r_*$. 

The interior of a BH is often described in the geometric optics limit, in which GWs behave as massless particles which follow null geodesics \cite{Misner:1973prb,Cardoso:2008bp}. The waves are described by Eq. \ref{RW_equation}, in which $\omega^2$ plays the role of an effective energy. If $\omega$ is sufficiently large, $V_\mathrm{RW}$ becomes negligible and one obtains the geometric optics limit, as illustrated in Fig. \ref{fig:potential}(b). Restoring the time derivative, Eq. \ref{RW_equation} becomes
\begin{equation}\label{wave_equation}
    \frac{\partial^2\psi}{\partial r_*^2}-\frac{\partial^2\psi}{\partial t^2}\sim  0,
\end{equation}
for asymptotically large $\omega$. This equation is solved by any function of $t\pm r_*$. In particular, it is solved by $\delta\left(t+r_*\right)$ and $\delta\left(t-r_*\right)$, which correspond to pointlike massless particles following null geodesics.

In this work, we are interested in wave behavior which is not within the geometric optics limit. By Eq. \eqref{RW_equation}, the internal minimum for axial gravitational perturbations can sustain bound states, each with negative effective energy $\omega^2$, as shown in the first row of Fig. \ref{fig:wavefunctions}. We focus on the simple case in which the effective energy is real. Thus, $\omega=i \omega_\mathrm{I}$ is purely imaginary, i.e., overdamped. Since only the square of $\omega$ appears in Eq. \eqref{RW_equation}, the same spatial solution is obtained regardless of the sign of $\omega_\mathrm{I}$. This sign is determined by the ingoing boundary condition at the horizon. We consider the region of $r_*$ to the left of the potential minimum in Fig. \ref{fig:potential}(b), where $V_\mathrm{RW}$ is negligible in Eq. \eqref{RW_equation}. The wavefunction $\psi$ should decay exponentially there for decreasing $r_*$, so $\psi\sim\mathrm{exp}(|\omega_\mathrm{I}| r_*)$. Furthermore, the RW master function is given by $\Psi(t,r)= \mathrm{exp}(-i\omega t) \psi(r) = \mathrm{exp}(\omega_\mathrm{I} t) \psi(r)$, where $t$ is the Schwarzschild time coordinate. When $\omega_\mathrm{I}>0$, then $\Psi\sim\mathrm{exp}(\omega_\mathrm{I} v)$, where $v\equiv t+r_*$ the ingoing null coordinate. Since $\Psi$ is a function of $v$, it is ingoing at the horizon, which is physically required due to the fact that the event horizon is a one-way hypersurface where anything that reaches it inevitably fall into the BH. A similar calculation for $\omega_\mathrm{I}<0$ shows that $\Psi\sim\mathrm{exp}(\omega_\mathrm{I} u)$ where $u\equiv t-r_*$ is the outgoing null coordinate. There is no outgoing information from the BH interior whatsoever, so we must use $\omega_\mathrm{I}>0$ to satisfy the boundary condition at the event horizon.

On the other hand, since $r_*\sim-r^2/4M$ as $r\to 0$, the potential \eqref{eq4} diverges as $3/4r_*^2$ close to the singularity. In this limit, Eq. \eqref{RW_equation} has a bound analytic solution of the form $\psi\sim(-r_* )^{3/2}$. In summary, the internal boundary conditions for axial perturbations are
\begin{equation}
	\label{internal_BCs}
	\psi \sim
	\left\{
	\begin{array}{lcl}
		(-r_* )^{3/2},\qquad\qquad\ r_*\to0, \\
		&
		&
		\\
		e^{\omega_\mathrm{I} r_*},\qquad\qquad\quad\,\,\, r_*\to -\infty.
	\end{array}
	\right.
\end{equation}
We find the OQBSs by numerically solving Eq. \eqref{RW_equation} with the boundary conditions \eqref{internal_BCs}, in order to obtain the bound states of the axial effective potential in Eq. \eqref{eq4}. We choose $\partial\psi⁄\partial r_*=1$ at some small starting value of $r_*$ in the range $10^{-5}<-r_*<3\times 10^{-4}$, and determine $\psi$ at this point via the $(-r_*)^{3/2}$ form of the boundary condition. Equation \eqref{RW_equation} is then integrated numerically from the starting point to some final negative value of $r_*$ which is chosen to be far enough from zero so that $\psi$ has decayed essentially to zero, as required from the boundary condition at the event horizon \eqref{internal_BCs} for OQBSs. The frequency $\omega_\mathrm{I}$ is optimized to minimize $\psi^2$ at the final value of $r_*$ ($\psi$ is real). For the sake of quantifying the numerical noise in our scheme, the process is repeated 30 times for equally-spaced starting values of $-r_*\in[10^{-5},3\times 10^{-4}]$.

\section{Results}

\begin{figure*}
    \centering
    \includegraphics[scale=0.53]{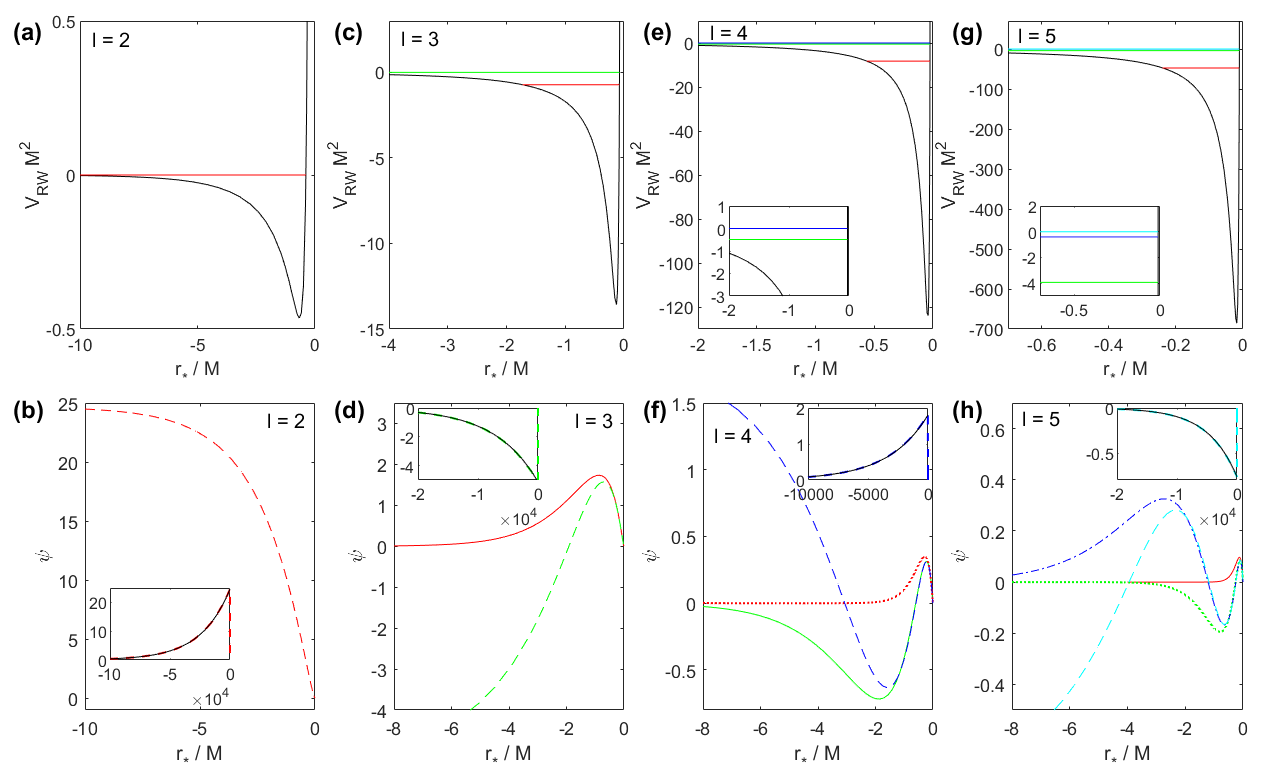}
    \caption{\emph{Top row} The overdamped gravitational modes for $\ell \leq 5$. (a), (c), (e) and (g) with respect to the effective potentials for $\ell = 2,\, 3,\, 4$ and $5$, respectively. The horizontal colored lines indicate the effective energy $-\omega_\mathrm{I}^2$ of each energy mode $n$. Insets are enlargements of the region close to zero effective potential. \emph{Bottom row} (b), (d), (f) and (h) show the wavefunctions for $\ell = 2,\, 3,\, 4$ and $5$, respectively. The color of each wavefunction matches the color of the energy state in the panel above. The number of nodes is equal to the number of the energy level $n$. The dashed curves corresponds to the smallest value of $|\omega_\mathrm{I}|$. Insets show the long range behavior of the dashed curves. The solid curves in the insets show the corresponding exponential dependence $\mathrm{exp}(\omega_\mathrm{I} r_* )$ of the smallest $|\omega_\mathrm{I}|$.}
    \label{fig:wavefunctions}
\end{figure*}

The upper row of Fig. \ref{fig:wavefunctions} depicts the OQBS levels which are supported by the effective potential for each $\ell$. The values of $-\omega_\mathrm{I}^2$ are indicated by colored horizontal lines. It is seen that larger choices of $\ell$ lead to deeper effective potential wells which support more energy states. The average and standard deviation of the resulting OQBSs of $\omega_\mathrm{I}$ are given in Table \ref{tab:table1}. The lower row of Fig. \ref{fig:wavefunctions} shows the eigenfunctions corresponding to each energy level. The number corresponding to the energy level of each OQBS is equal to the number of nodes in the wavefunction. The lowest, most negative (largest in absolute value) energy state has zero nodes, so it corresponds to the ground state $n=0$. It is obvious from Fig. \ref{fig:wavefunctions} that excited states (shown with different colors) have more nodes, and thus define the excited OQBSs $n=1,\,2,\,3$, and so on. The overall behavior of the eigenfunctions shown in Fig. \ref{fig:wavefunctions} satisfy the boundary conditions \eqref{internal_BCs}, i.e., they tend to zero by a power law close to the singularity and decay exponentially at the event horizon, as a function of $r_*$. Nevertheless, in order to fully grasp the physical interpretation of these modes, we switch to Kruskal-Szekeres coordinates, which unveils that the full wavefunction encounters the event horizon.

\begin{table}[t]
    \centering
    \resizebox{\columnwidth}{!}{
    \begin{tabular}{c|c|c|c|c}
         \hline\hline
         & $\ell=2$ & 3 & 4 & 5\\
         \hline
        $n=0$ & $7.7(6)\times 10^{-5}$	&
        1.70573(2) & 5.74623(7) & 13.7683(3) \\
        1 & -- & $2.37(9)\times 10^{-4}$ & 1.39464(3) & 3.9691(1)\\
        2 & -- & -- & $3.7(6)\times 10^{-4}$ & 1.27485(5) \\
        3 & -- & -- & -- & $5.7(4)\times 10^{-4}$\\
        \hline\hline
    \end{tabular}}
    \caption{The frequencies of the purely imaginary quasi-bound states, up to $\ell=5$. The values are the $\omega_\mathrm{I}$ for each $\ell$, in units of $1/2M$, where the wavefunction has $n$ nodes. The values in parentheses indicate the one-standard deviation uncertainty in the last digit, due to numerical noise.}
    \label{tab:table1}
\end{table}

\begin{figure}[t]
\vspace*{\fill}\begin{adjustbox} {angle=0,center,nofloat=table} 
 \includegraphics[width=\columnwidth]{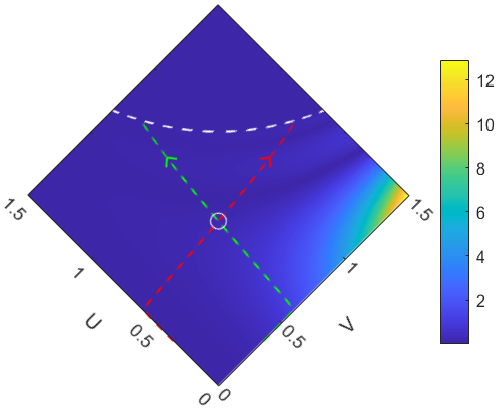}
\end{adjustbox}
\vspace*{\fill}
    \caption{An OQBS in Kruskal-Szekeres coordinates. The absolute value of $\Psi(t,r)$ for $\ell=5$, $n=2$ is shown. The diagram corresponds to quadrant II in a Kruskal-Szekeres diagram. The $U=0$ ($V=0$) line is the future (past) horizon. The dashed $UV=1$ curve is the singularity. The red and green curves are examples of timelike geodesic observers. Their world lines happen to cross at the white circle.}
    \label{fig:UVdiagram}
\end{figure}

To obtain a definitive picture on how an OQBS behaves at each point in spacetime, we transform the BH metric in study from Schwarzschild coordinates $(t,\,r)$ to Kruskal-Szekeres coordinates $(U,\,V)$ \cite{Kruskal:1959vx,Szekeres:1960gm,Israel:1966zz}. These coordinates are well-behaved everywhere outside the curvature singularity at $r_*=0$. Fig. \ref{fig:UVdiagram} shows an OQBS in Kruskal-Szekeres coordinates, where deep blue indicates zero magnitude while brighter colors indicate larger values. We observe that the wavefunction vanishes at the singularity (dashed white curve at $UV=1$) and at the past event horizon (deep blue along $V=0$). We also find that the wavefunction fully encounters the future event horizon (brighter colors appear at $U=0$). Thus, the state is formally quasibound, i.e., it decays (to zero) at the singularity but it is free (non-zero and finite) at the event horizon.

\begin{figure}[t]
 \includegraphics[width=\columnwidth]{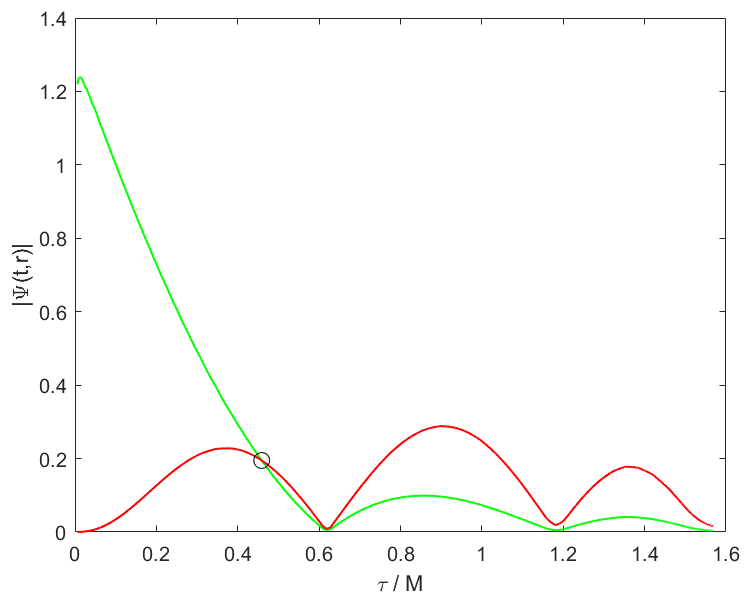}
    \caption{An OQBS as seen by geodesic observers. The red and green curves are the values from Fig. \ref{fig:UVdiagram} along the red and green world lines, as functions of each observer's proper time. The two curves are equal at the black circle, which corresponds to the white circle in Fig. \ref{fig:UVdiagram}.}
    \label{fig:psiOftau}
\end{figure}

In order to determine the evolution of the mode according to a given observer, one should determine the path of the observer through the spacetime of Fig. \ref{fig:UVdiagram}. This approach avoids issues of Schwarzschild coordinates, such as the reversal of time and space inside the BH \cite{Fiziev:2006tx}. The red and green curves in Fig. \ref{fig:UVdiagram} show examples of two such observers on timelike geodesics, each with almost constant $U$ or $V$. Thus, these two world lines almost form a light cone. Fig. \ref{fig:psiOftau} shows the values from Fig. \ref{fig:UVdiagram} along the red and green world lines. The values are equal at the black circle, which corresponds to the white circle in Fig. \ref{fig:UVdiagram}, where the world lines happen to cross (see Appendix \ref{AppA1} for more details). Figure \ref{fig:psiOftau} ultimately demonstrates that the OQBS decays with proper time for both observers as they approach the singularity, in accord with the boundary conditions. Furthermore, oscillations are seen in Fig. \ref{fig:psiOftau} as a function of proper time, which reflect the oscillations as a function of $r_*$ seen in Fig. \ref{fig:wavefunctions}, while there are no oscillations as a function of the coordinate time $t$. This association of proper time with $r_*$ is expected inside a black hole, where the roles of space and time are interchanged.

It is interesting to ask whether the OQBSs produce finite metric perturbations at the singularity. In the geometric optics limit illustrated in Fig. \ref{fig:potential}(b) \cite{Misner:1973prb,Cardoso:2008bp}, $V_\mathrm{RW}$ is negligible compared to $\omega^2$ in Eq. \eqref{RW_equation} (except near the singularity), and the gravitational perturbations can come very close to the singularity, although they would not reach it due to the infinite potential barrier at $r=0$. The OQBS wavefunction vanishes at the singularity like $\psi\sim(-r_* )^{3/2}$, and the radial part of the axial metric perturbation is related to the quantity $r\,\psi(r)\sim r_*^2$ \cite{Chandrasekhar:1985kt}, so it tends to zero at $r=0$ (see also Eq. (B.21) in \cite{Lousto:2005xu}). Therefore, the OQBSs produce vanishing metric perturbations at the singularity $r=0$.

Finally, we point out that the isospectrality property between axial and polar metric perturbations seems to be broken in the interior. It certainly does not hold for QBSs of massive fields at the exterior of Schwarzschild BHs \cite{Rosa:2011my,Brito:2013wya}. By breaking down the isospectrality problem into the exterior and interior of a Schwarzschild BH we conclude the following. In the exterior, axial and polar gravitational QNMs are isospectral. It has been shown analytically that the Regge-Wheeler (axial sector) and Zerilli (polar sector) equations are related via a transformation \cite{Chandrasekhar:1975zza,Chandrasekhar:1985kt}, that was later associated with the Darboux transformation \cite{Darboux:1999tra} and its generalized framework \cite{Heading:1977,Anderson:1991,Leung:1999,Brink:2000}, rendering the origin of the isospectrality more mathematically sound \cite{Glampedakis:2017rar}. In addition to the perturbation invariance under the Darboux transformation, the boundary conditions make QNMs isospectral. In particular, the purely ingoing-wave condition at the event horizon and purely-outgoing wave condition at infinity are preserved under the Darboux transformation, therefore leading to isospectrality. The exterior spectral problem essentially shows that isospectrality is not only an outcome of the wave equation, but is also intrinsically and significantly tied to the boundary conditions. As we show in Appendix \ref{AppA}, the regularity conditions at $r=0$ are not preserved under Darboux transformations and therefore the isospectrality of polar and axial metric perturbations might be broken in the BH interior. This is similar to what happens in BHs in anti-de Sitter spacetime \cite{Berti:2003ud} for example, or even in exotic compact objects \cite{Saketh:2024ojw}, that have different boundary conditions from the usual QNM ones at infinity and at the (would-be) horizon, and therefore do not satisfy any isospectrality relation.

Moving the discussion to the interior, and keeping in mind that the boundary conditions in Eq. \eqref{internal_BCs} are not invariant under the Darboux transformation (at least the regularity condition close to the singularity is not satisfied, see Appendix \ref{AppA} for a detailed proof), we can assume that the interior axial OQBSs found here, may not satisfy the isospectrality relation with respect to the interior modes of polar perturbations.

\section{Discussion} 

The interior of a Schwarzschild BH is traditionally regarded as a region where all physical fields are inexorably driven into the singularity. In this work, we have demonstrated that axial gravitational perturbations in Schwarzschild BH interiors can give rise to OQBSs that linger temporarily, before decaying, as they drift towards the central singularity. Specifically, using Kruskal–Szekeres coordinates, we identified and analyzed these modes, showing that their Regge–Wheeler master wavefunctions remain finite at the future event horizon while decaying before the singularity. This result runs counter to the common assumption that all disturbances inside the BH are ineluctably drawn directly into the singularity, pointing instead to a richer and more nuanced landscape of the dynamics within BH interiors.

In addition to studying a spectral problem in the interior of a Schwarzschild BH, we have found that axial metric perturbations can hover transiently between the horizon and the singularity via these OQBSs, that retain their regularity up to $r=0$. The results of this work apply to axial gravitational perturbations, only. For scalar or polar gravitational perturbations, the potential approaches $-\infty$ like $-1/4r_*^2$ at the central singularity, which implies that the wavefunction vanishes like $\psi\sim(-r_* )^{1/2}$ there. The vanishing of the wavefunction seems qualitatively similar to the case of axial perturbations, but the unbounded potential that diverges to $-\infty$, together with the lack of a potential well, presents a rather different problem from the axial case. From a numerical point of view, the polar case seems unstable against the formation of short wavelength oscillations in the large negative potential.

An important reminder here is the fact that all physically-meaningful conclusions have been drawn from the wave dynamics of axial gravitational perturbations in the BH interior analyzed in Kruskal-Szekeres, instead of standard Schwarzschild coordinates. That being said, the OQBSs might be considered as ``bound states in time'' in a traditional Schwarzschild-coordinate perspective, since in the BH interior the conventional coordinate $r$, and thus $r_*$, is timelike. It is rather interesting, though, that the concept of such bound states has already been considered in other physical settings. For instance, Ref. \cite{Schiller:2024} presents an analogous phenomenon in the context of Maxwell’s equations in a time-varying, spatially homogeneous medium. The equations used in \cite{Schiller:2024} share a very similar structure with those studied here, with the respective effective potential arising from a time-dependent refractive index. This parallel reinforces the idea that such solutions can have a variety of meaningful physical interpretations.

Even though our analysis lies purely at the level of classical perturbation theory, there is a plurality of applications regarding quantum effects, semi-classical and eventually quantum gravity in BH interiors \cite{Russo:1993br,Nomura:2014woa}. Here, we have obtained interior purely imaginary resonances. If appropriate regularity conditions are satisfied by the resonance wavefunctions, those may reflect decay rates or spatial modes relevant to infalling observers, semi-classical stress tensors, or even effective field theory on curved spacetime. Of course, these resonances do not couple in any physical way to the exterior, and resulting observables, but they might influence internal wave dynamics and late-time tails \cite{Iliesiu:2021ari}, how semi-classical fields diverge or remain finite near the singularity and possible holographic duals or quantum-gravity boundary conditions \cite{Anegawa:2025tio}. Thus, these resonances are physically meaningful in BH interior physics, especially if we consider how fields behave near the singularity or how quantum corrections might modify it.

To further elaborate, we may utilize interior resonances, such as those found here, in order to study the expectation value $\langle T_{\mu\nu}\rangle$ of the quantum stress-energy tensor in a classical spacetime background. The behavior of fields near the singularity controls whether $\langle T_{\mu\nu}\rangle$ becomes divergent or can be regularized \cite{Alesci:2019pbs}. Unlike full quantum gravity, the study of wave equations in BH interiors is analytically and numerically accessible. Interior resonances give a concrete playground for testing ideas like singularity resolution \cite{Modesto:2006mx} and quantum chaos \cite{Addazi:2015gna,Maldacena:2015waa,Turiaci:2016cvo,Perlmutter:2016pkf}. If discrete interior modes dominate the structure of the field near the singularity, they may effectively ``organize'' or ``tame'' that divergence \cite{Modesto:2004xx,Boehmer:2008fz,Gambini:2013ooa}. This is completely analogous to how bound states determine spectral properties in quantum mechanics. 

In quantum gravity, especially in path integral or Euclidean approaches, boundary conditions at singularities or at finite radii are required \cite{Chua:2023srl}. Interior resonances could offer a natural selection mechanism for such boundary conditions, e.g., only allowing fields that are smooth or decay in a certain way at $r = 0$ \cite{Modesto:2005zm,Ashtekar:2005qt}. These can even be viewed as ``inner quantization conditions'' \cite{Wei:2009yj}, reminiscent of the Bohr-Sommerfeld conditions in semi-classical quantum mechanics \cite{Festuccia:2008zx}. In speculative models where BHs have dual descriptions \cite{Maldacena:1997re,Maldacena:2024qlf}, e.g., fuzzballs \cite{Mathur:2005zp,Bianchi:2018kzy,Bianchi:2020miz}, firewalls \cite{Kaplan:2018dqx}, and holographic duals \cite{Lunin:2001jy,Mathur:2012dxa,Papadodimas:2013wnh}, the interior degrees of freedom should encode information about microstates \cite{Ikeda:2021uvc}. Interior resonant modes might act like a spectral fingerprint of interior microstructure, even if the actual microstates are unknown. So, even though these modes are not directly observable from outside the BH, they can influence the behavior of semi-classical fields, provide natural regularization or extension criteria, guide the construction of effective theories with modified interiors, or be incorporated into quantum corrections, e.g., via mode sum regularization \cite{Marolf:2018ldl}.

\begin{acknowledgements}
We thank Valentin Boyanov, Vitor Cardoso, Nicola Franchini, David Hilditch, Kostas D. Kokkotas, Adrien Kuntz, Jorge Patiño, Hannes Rüter and Sebastian V\"olkel for helpful discussions. 
This work was supported by the Israel Science Foundation, grant 531/22.
K.D. acknowledges financial support provided by FCT – Fundação para a Ciência e a Tecnologia, I.P., under the Scientific Employment Stimulus – Individual Call – Grant No. 2023.07417.CEECIND/CP2830/CT0008.
R.B. acknowledges financial support provided by FCT – Fundação para a Ciência e a Tecnologia, I.P., through the ERC-Portugal program Project ``GravNewFields''.
K.D. and R.B. also acknowledge financial support provided by the Fundação para a Ciência e Tecnologia (FCT), Portugal, for the financial support to the Center for Astrophysics and Gravitation (CENTRA/IST/ULisboa) through grant No. UID/PRR/00099/2025 and grant No. UID/00099/2025.
This project has received funding from the European Union’s Horizon MSCA-2022 research and innovation programme “Einstein Waves” under grant agreement No. 101131233.
\end{acknowledgements}

\appendix

\section{Radial timelike geodesics\label{AppA1}}

A timelike geodesic is given by the values of $r$ and $t$ at each proper time $\tau$. For a radially infalling observer the relation between $r$ and $\tau$ is given by \cite{Foster:2006}
\begin{align}\nonumber
\tau&=\frac{r_0^{3/2}}{(2M)^{1/2}}\left[\frac{\pi}{2}-\mathrm{arcsin}\left[\left(\frac{r}{r_0}\right)^{1/2}\right]\right.\\
&\left.+\left(\frac{r}{r_0}\right)^{1/2}\left(1-\frac{r}{r_0}\right)^{1/2}\right]
\end{align}
where the massive object is at rest at $r=r_0$ and $\tau=0$. For the particular geodesics shown in Fig. \ref{fig:UVdiagram}, $r_0=4.54M$. The $t$ coordinate is then determined by integrating
\begin{equation}
dt=\frac{k}{1-2M/r}d\tau
\end{equation}
where $k^2=1-2M/r_0$. The start time of each geodesic is adjusted such that $t=-11.5M$ at $\tau=0$ for the green geodesic, and $t=11.5M$ at $\tau=0$ for the red geodesic. In Fig. \ref{fig:psiOftau}, the origin of $\tau$ is adjusted so $\tau=0$ corresponds to the moment that the observer is at the horizon.

\section{Isospectrality breaking in Schwarzschild black-hole interiors}\label{AppA}

Here, we analyze the Regge-Wheeler (axial) and Zerilli (polar) master equations near the curvature singularity $r=0$ in the Schwarzschild BH interior ($r<2M$). Not only the Regge-Wheeler, but also the Zerilli master equation, take the Schr\"odinger-like form
\begin{equation*}
\frac{d^2\psi}{dr_*^2} + \big(\omega^2 - V\big)\psi = 0,
\end{equation*}
as discussed in Sec. \ref{SecII}. The Regge-Wheeler potential reads
\begin{equation*}
V_{\rm RW}(r) = f\!\left( \frac{\ell(\ell+1)}{r^2} - \frac{6M}{r^3} \right),
\end{equation*}
while the Zerilli potential has the form 
\begin{equation*}
V_{\rm Z}(r) = \frac{2f}{r^3}\,
\frac{9M^3 + 9M^2\lambda r + 3M\lambda^2 r^2 + \lambda^2(\lambda+1)r^3}{(3M+\lambda r)^2},
\end{equation*}
with $\lambda = \tfrac{1}{2}(\ell-1)(\ell+2)$. A quick glance reveals that the two potentials differ dramatically. Near $r=0$, we have $f(r\rightarrow 0)\sim -2M/r$ and $f'(r\rightarrow 0)\sim 2M/r^2$. The leading-order terms of the potentials as $r\rightarrow 0$ are
\begin{equation}
V_{\rm RW}(r) \;\sim\; \frac{12M^2}{r^4}, 
\qquad
V_{\rm Z}(r) \;\sim\; -\,\frac{4M^2}{r^4}.
\end{equation}
The crucial point to note is the opposite signs and different coefficients of the $r^{-4}$ terms.

\subsection{Frobenius equation and indices at $r=0$}

Writing the master equation in terms of $r$, i.e. expanding the second derivative with respect to the tortoise coordinate $r_*$, we get
\begin{equation}\label{master_r}
f^2 \psi'' + f f' \psi' + (\omega^2 - V)\psi = 0,
\end{equation}
where a prime denotes differentiation with respect to $r$. 

Here, we use the Frobenius method in order to find the indicial solutions of the resulting quadratic equation derived from the coefficients of a power series solution to the second-order ordinary differential equation in \eqref{master_r} in the vicinity of the regular singular point $r=0$, i.e. the singularity. By substituting $\psi \sim r^p$ in \eqref{master_r}, and taking into account that $\omega^2\psi$ is sub-leading to $V\psi$ near $r=0$, we find the leading contribution as
\begin{align}\nonumber
f^2 \psi'' + f f' \psi'-V\psi&=0 \Rightarrow\\ 4M^2\,(p^2-2p)\, r^{p-4}-V r^p&=0.\label{Frobenius}  
\end{align}

For the Regge-Wheeler case, we obtain
\begin{align}\nonumber
4M^2(p^2-2p)\,r^{p-4} - \frac{12M^2}{r^4}\, r^p &= 0 \Rightarrow\\ \label{indicial_RW}
p^2 - 2p - 3 &= 0,
\end{align}
where the solutions to the indicial equation \eqref{indicial_RW} are $p=-1,\,3$. Thus, according to the Frobenius theorem, one local branch near the singularity has the form $\psi_\textrm{RW}\sim r^{-1}$, which is singular, while the other branch has the form $\psi_\textrm{RW}\sim r^{3}$, which is regular.

For the Zerilli case, we similarly have
\begin{align}\nonumber
4M^2(p^2-2p)\,r^{p-4} - \left(-\frac{4M^2}{r^4}\right)r^p &= 0 \Rightarrow\\ \label{indicial_Z}
p^2 - 2p + 1 &= 0,
\end{align}
with indicial solution the double root $p=1$. Thus, since the root here is degenerate, the Frobenius theorem states that the local behavior should have the form, $\psi_{\rm Z} \sim r$ and the form $\psi_{\rm Z} \sim r\ln r$, which has a weaker differentiability (its derivative diverges logarithmically) but it is still $\mathcal{O}(1)$ as $r\rightarrow 0$. For this case, both solutions are regular, i.e. vanishing at $r=0$.

\subsection{Behavior under the Chandrasekhar-Darboux map}

Chandrasekhar showed that the Zerilli and Regge-Wheeler equations are related via a transformation \cite{Chandrasekhar:1985kt}. By defining the transformation operators
\begin{align}
\mathcal{D} &= \frac{d}{dr_*} + W(r), \\
\tilde{\mathcal{D}} &= -\frac{d}{dr_*} + W(r),
\end{align}
where $W(r)$ is constructed from the background geometry and $\ell$, the transformation should satisfy the intertwining relation
\begin{equation}
\mathcal{D}\left(\frac{d^2}{dr_*^2} + \omega^2 - V_{\text{RW}}\right) = \left(\frac{d^2}{dr_*^2} + \omega^2 - V_{\text{Z}}\right)\mathcal{D}.
\end{equation}
This essentially means that if $\psi_\textrm{RW}$ solves the Regge-Wheeler equation, then $\psi_\textrm{Z} = \mathcal{D}\,\psi_\textrm{RW}$ solves the Zerilli equation with the same frequency $\omega$. The inverse transformation is given by $\psi_\textrm{RW} = \tilde{\mathcal{D}}\,\psi_\textrm{Z}$.

The Chandrasekhar-Darboux relation has an explicit form
\begin{align}
\psi_{\rm Z} &= \frac{\lambda(\lambda+1)r^2 + 3\lambda M r + 6M^2}{r(\lambda r + 3M)} \, \psi_{\rm RW}\nonumber
\\&+ \frac{3M f}{\lambda r + 3M}\, \frac{d\psi_{\rm RW}}{dr_*}, \label{map1} \\\nonumber
\psi_{\rm RW} &= \frac{\lambda(\lambda+1)r^2 - 3\lambda M r + 6M^2}{r(\lambda r + 3M)} \, \psi_{\rm Z}
\\ &- \frac{3M f}{\lambda r + 3M}\, \frac{d\psi_{\rm Z}}{dr_*}. \label{map2}
\end{align}
Equations \eqref{map1} and \eqref{map2} provide the mapping between the axial and polar master equations. Note that the only singular behavior in Eqs. \eqref{map1} and \eqref{map2} results from the tortoise coordinate derivative
\begin{equation}
\frac{d}{dr_*} = f(r)\frac{d}{dr} \sim -\frac{2M}{r}\frac{d}{dr}
\quad (r\to 0).    
\end{equation}
Thus, the possible fate of isospectrality can be decided from the requirement of regularity in the vicinity of $r=0$. If $\psi_{\rm RW} \sim r^3$, then
\begin{equation}
\frac{d\psi_{\rm RW}}{dr_*} \sim -\frac{2M}{r} 3r^2 \sim -6M r,
\end{equation}
so from \eqref{map1}, $\psi_{\rm Z} \sim r$, i.e. the transformation from the Regge-Wheeler to Zerilli wavefunctions is regular. By performing the same test for the inverse transformation, provided in Eq. \eqref{map2}, we find that if $\psi_{\rm Z} \sim r$, then
\begin{equation}
\frac{d\Psi_{\rm Z}}{dr_*} \sim -\frac{2M}{r}.
\end{equation}
Therefore Eq. \eqref{map2} yields $\psi_{\rm RW} \sim r^{-1}$, i.e. the inverse transformation leads to the singular axial branch. Hence, the Chandrasekhar-Darboux transformation does not preserve the regularity boundary condition at the vicinity of the singularity ($r=0$) in both directions, meaning that
\begin{equation}
\psi_{\rm RW}^{\rm reg} \mapsto \psi_{\rm Z}^{\rm reg}, 
\,\,\,\,\qquad
\psi_{\rm Z}^{\rm reg} \mapsto \psi_{\rm RW}^{\rm sing}.
\end{equation}
Therefore, although the Regge-Wheeler and Zerilli equations are Darboux-related, their boundary-value problems are may not be isospectral in the BH interior. Hence, axial and polar modes inside Schwarzschild BHs may not share the same spectra.

\bibliography{biblio}

\end{document}